\documentclass[amsmath,superscriptaddress,amssymb,twocolumn]{revtex4}
\def\beqa{\begin{eqnarray}}
\def\eeqa{\end{eqnarray}}
\def\beqn{\begin{equation}}
\def\eeqn{\end{equation}}

\def\PN{\Phi_N}
\def\PNa{{\Phi_a}}
\def\PP{{\Phi_P}}
\def\Pst{\phi}

\def\lr{{\bar r}}        
\def\gb{{\bar g}}        %
\def\PbN{\delta {\bar \Phi}_N}   
\def\PbP{\delta {\bar \Phi}_P}

\def\k{k}                  
\def\ks{\mathbf{\k}}
\def\s{s}                  

\def\M{M}                  
\def\GN{G_N}

\def\GP{G_P}

\def\Gts{{\tilde G^0}}
\def\Gtt{{\tilde G^1}}
\def\GtN{{\tilde G_N}}
\def\GtNa{{\tilde G_a}}
\def\GtP{{\tilde G_P}}

\def\stand#1{\left[#1\right]_\mathrm{st}}

\begin{document}

\title{LONG RANGE GRAVITY TESTS AND THE PIONEER ANOMALY}

\author{Serge Reynaud}

\address{Laboratoire Kastler Brossel\footnote{CNRS, ENS, UPMC}, 
Universit\'e Pierre et Marie Curie, case 74, \\
Campus Jussieu, F75252 Paris cedex 05 \\
reynaud@spectro.jussieu.fr}

\author{Marc-Thierry Jaekel}

\address{Laboratoire de Physique Th\'eorique\footnote{
CNRS, ENS, UPMC}, Ecole Normale Sup\'{e}rieure, \\
24 rue Lhomond, F75231 Paris cedex 05 \\
jaekel@lpt.ens.fr}

\begin{abstract}
Experimental tests of gravity performed in the solar system show a good agreement 
with general relativity. The latter is however challenged by the Pioneer anomaly 
which might be pointing at some modification of gravity law at ranges of the order 
of the size of the solar system. As this question could be related to the puzzles 
of ``dark matter'' or ``dark energy'', it is important to test it with care. 
There exist metric extensions of general relativity which preserve the well verified 
equivalence principle while possibly changing the metric solution in the solar system. 
Such extensions have the capability to preserve compatibility with existing gravity 
tests while opening free space for the Pioneer anomaly. They constitute arguments 
for new mission designs and new space technologies as well as for having a new look 
at data of already performed experiments.
\end{abstract}
\maketitle

\section{Introduction}

The commonly heard assertion that gravity tests show a good agreement with general 
relativity (GR) has to be understood as a set of more detailed 
statements \cite{Will93,Will01}.
It first implies that the gravitational field may be identified with the metric tensor 
$g_{\mu\nu}$ in a Riemannian space-time, as a consequence of the fact that the 
equivalence principle is one of the most accurately verified properties of nature.
It then means that this metric tensor appears to have a form close to that predicted 
by GR, as shown by the confrontations of observations with the family of more general 
PPN solutions. This second statement can be put under the alternative form of a good
agreement of the gravity force law with the prediction of GR, deviations being 
predicted by unification models but not observed 
to date \cite{Fischbach98,Adelberger03}.

Besides these successes, GR is challenged by observations performed at various scales. 
{}First, anomalies are known to affect the rotation curves of galaxies.
They are commonly accounted for by introducing ``dark matter'' to reproduce these 
curves \cite{Aguirre,Riess}. 
{}Further anomalies have been detected more recently in the relation between redshifts 
and luminosities, showing an acceleration of cosmic expansion. They are usually 
interpreted as due to the presence of some 
``dark energy'' \cite{Perlmutter99apj,Perlmutter99prl}.
Both components of the ``dark side'' of the universe have no known origin and are 
not observed through other means than the gravitational anomalies they have been 
designed to cure. 
As long as this situation is lasting, the related anomalies may as well be interpreted 
as long range deviations from GR \cite{Sanders02,Lue04,Turner04}.

The Pioneer anomaly constitutes a new piece of information in this puzzling context,
which might already reveal an anomalous behaviour of gravity at scales 
of the order of the size of the solar system \cite{Anderson98,Anderson02}.
Though a number of mechanisms have been considered to this 
aim \cite{Anderson02b,Anderson03,Nieto04,Turyshev04,Nieto05a,Nieto05},
the anomaly has escaped up to now all attempts of explanation as a systematic
effect generated  by the spacecraft itself or its environment.
The importance of the Pioneer anomaly for space navigation already justifies it to be 
submitted to further scrutiny.
Meanwhile its potential impact on fundamental physics cannot be underestimated, 
since the possibility exists that the Pioneer anomaly be the first hint of a long 
range modification of gravity law \cite{Bertolami04,JR05mpl,JR05cqg,JR06cqg,Brownstein06,Lammerzahl06,JR06submitted}.

These questions are reviewed in the present paper, with the emphasis put on the key 
issue of the compatibility of the Pioneer anomaly with other gravity tests. 

\section{Gravity tests in the solar system}

General Relativity (GR) provides us with an excellent description of gravitational 
phenomena in the solar system. In order to discuss the meaning of this common 
statement, we first recall the basic features of this theoretical description 
and then briefly review the experimental evidences supporting it.

In order to apply the principle of relativity to accelerated motions, Einstein  
introduced what is now called the 
equivalence principle \cite{Einstein07,Einstein11}.
A weak form of this principle is expressed by the universality of free fall, 
a central property of the theory of gravitation since Galileo and Newton 
which acquires with Einstein a geometrical significance,  
gravitation fields being identified with the metric tensor $g_{\mu\nu}$
in a Riemannian space-time.
Ideal atomic clocks measure the proper time $\int ds$ along their trajectory 
in space-time with $ds^2\equiv g_{\mu\nu}dx^\mu dx^\nu$.
Meanwhile freely falling motions are the geodesics of this Riemannian space-time,
that is also the curves which extremize the integral $\int ds$. 

The equivalence principle is one of the best ever tested properties of nature.
Potential violations are usually parametrized by a relative difference $\eta$ 
in the accelerations $a_1$ and $a_2$ undergone by two test bodies of different
compositions in free fall at the same location and with the same velocity. 
Modern experiments constrain the parameter $\eta$ to stay below the $10^{-12}$ level.
These experiments test the principle at distances ranging from the
millimeter in laboratory experiments (Ref.~\cite{Adelberger03} and references in)
to the sizes of Earth-Moon \cite{Williams96} or Sun-Mars orbit \cite{Hellings83}.

The geometrical interpretation is the very core of GR, but it is not sufficient 
to fix the latter theory. In order to do that, it is necessary to write also 
the equations determining the metric tensor from the distribution of energy and 
momentum in space-time or, in other words, to fix the form of the coupling between 
curvature and stress tensors. 
Among the curvature tensors available in Riemannian geometry, 
the Einstein tensor $E _{\mu\nu}\equiv R _{\mu\nu} - {1\over2} g _{\mu\nu} R$ is
defined from the Ricci tensor $R _{\mu\nu}$ and scalar curvature $R$ so that
it has a null covariant divergence $D^\mu E_{\mu\nu}\equiv0$.
This geometrical property has to be compared with the physical property 
$D^\mu T_{\mu\nu}\equiv0$ 
which expresses conservation of energy and momentum as the condition of null 
divergence of the stress tensor $ T _{\mu\nu}$.
Note that the latter relation is a necessary and sufficient condition 
for motions of test masses to follow geodesics.

GR corresponds to a simple proportionality relation between the two tensors $E _{\mu\nu}$
and $T _{\mu\nu}$, the constant being determined from the Newton gravitation 
constant $\GN$ and the velocity of light $c$,
\beqa
\label{GR_gravitation_law}
E _{\mu\nu} = {8\pi\GN \over c^4} T _{\mu\nu}
\eeqa
This Einstein-Hilbert equation \cite{Einstein15,Hilbert,Einstein16} is tested 
through comparisons of its predictions with observations or experiments. 
To this aim, the metric tensor in the solar system is first deduced
by solving (\ref{GR_gravitation_law}). 
In the simple case where the gravity source, \textit{i.e.} the Sun, 
is described as a point-like motion-less mass $\M$,
the metric can be written as an expansion in terms of the Newton potential $\Pst$
\beqa
\label{GR_metric}
d\s^2 &=&  g _{00}  c ^2 d t ^2 +  g _{rr} \left( dr^2  +
 r ^2(d\theta^2 + {\rm \sin}^2\theta  d\varphi^2) \right)\\
\label{GR_solar_metric}
 g _{00} &=& 1+2\Pst+2\Pst^2+\ldots\quad, \quad
 g _{rr} = - 1+2\Pst+\ldots \nonumber\\
\Pst &\equiv& -{\kappa\over r}\quad, \quad 
\kappa \equiv {\GN\M\over c^2}\quad, \quad 
\left\vert\Pst\right\vert \ll 1\nonumber
\eeqa
Spherical coordinates have been used ($t$ and $r$ are time and radius,
$\theta$ and $\varphi$ colatitude and azimuth angles)
with the Eddington gauge convention of isotropic spatial coordinates.
$\kappa$ is the gravitational radius of the Sun $\sim 1.5$km.

GR is usually tested through its confrontation with the enlarged family of 
parametrized post-Newtonian (PPN) metric tensors
introduced by Eddington \cite{Eddington} and then developed by  
several physicists \cite{Robertson,Schiff66,Nordtvedt68,WillNordtvedt72} 
\beqa
\label{PPN_0}
 g _{00} = 1 + 2 \alpha \Pst + 2 \beta \Pst^2 + \ldots \quad,\quad 
 g _{rr} = -1 + 2 \gamma \Pst + \ldots &&
\eeqa
The three parameters $\alpha$, $\beta$ and $\gamma$ are constants, the 
first of which can be set to unity by fixing Newton constant $\GN$.
Within the PPN family, GR thus corresponds to $\gamma=\beta=1$.
Anomalies $\gamma-1$ or $\beta-1$ of these Eddington parameters affect 
motions, \textit{i.e.} the geodesics associated with the metric (\ref{PPN_0}),
and they can therefore be measured by comparing observations with predictions. 

Experiments now performed for more than four decades have led
to more and more constraining bounds on these anomalies. 
{}For example, Doppler ranging on Viking probes in the vicinity of 
Mars \cite{Hellings83} and deflection measurements using VLBI astrometry \cite{Shapiro04} 
or radar ranging on the Cassini probe \cite{Bertotti03} have given smaller and smaller
values of $\vert\gamma-1\vert$, with presently a bound of a few $10^{-5}$. 
Analysis of the precession of planet perihelions \cite{Talmadge88} and of 
the polarization by the Sun of the Moon orbit around the Earth \cite{LLR02} 
have led to determinations of linear superpositions of $\beta$ and $\gamma$, 
resulting now to $\vert\beta-1\vert$ smaller than a few $10^{-4}$.

An alternative manner to test GR has been to check the $r-$dependence of the Newton 
potential, that is also of the component $ g _{00}$ in (\ref{GR_solar_metric}).
Hypothetical modifications of its standard expression, predicted by unification models,
are usually parametrized in terms of an additional Yukawa potential depending on two 
parameters, the range $\lambda$ and the amplitude $\alpha $ measured with respect to 
Newton potential \cite{Fischbach98}.
The presence of such a Yukawa correction has been looked for on a large
range of distances.
The accuracy of short range tests has been recently improved, as gravity experiments 
were pushed to smaller distances \cite{Hoyle,Long,Chiaverini} and as Casimir forces, 
which become dominant at submillimeter range, were more satisfactorily taken into 
account \cite{Bordag01,Lambrecht02,Decca03,Chen04}.
On the other side of the distance range, long range tests of the Newton law are 
performed by following the motions of planets or probes in the solar system. 
They also show an agreement with GR with a good accuracy for ranges of the order 
of the Earth-Moon \cite{Williams96} or Sun-Mars 
distances \cite{Reasenberg79,Anderson96,Kolosnitsyn03}.
When the whole set of results is reported on a global figure (see Fig.~1 in
Ref.~\cite{JR04} reproduced thanks to a courtesy of Coy \textit{et al} \cite{Coy03}),
it appears that windows remain open for violations of the standard
form of Newton force law at short ranges, below the millimeter, as well as
long ones, of the order of or larger than the size of the solar system.

To sum up this discussion, tests of gravity confirm its metric interpretation
and provide strong evidence in favor of gravitation theory being very close to
GR. A few exceptions exist, 
among which notably the anomalous observations recorded on Pioneer probes.
We will see below that this contradiction between Pioneer observations
and other gravity tests may be cured in an extended framework, where
deviations from GR may show a scale dependence.
It is precisely the merit of Newton force law tests to shed light on this 
possibility of a scale dependence, with any specific experiment being only sensitive
to a given range of distances. 
The issue of scale dependence has to be considered with great attention,
especially in the context recalled in the Introduction where questions arise
about the validity of GR at galactic or cosmic scales.
As recalled in forthcoming sections, scale dependence is also a natural consequence 
of radiative corrections to GR to be taken into account.

\section{The Pioneer anomaly}

After the discussions of the previous section, it is clear that the gravity laws
have to be tested at all possible scales. It is of particular interest
to study the largest scales attainable by man made instruments, in an attempt
to bridge the gap between experiments made on Earth or in its vicinity and
the much larger galactic and cosmic scales. The best example of such a strategy
to this date is the NASA decision to extend Pioneer 10 \& 11 missions after their 
primary periods with the aim, among others, to test the laws of gravity at 
large heliocentric distances \cite{Fimmel77,Fimmel80}.
When considered as a gravity test, the extended Pioneer missions were the largest
scaled test ever carried out, and they failed to confirm the known laws of gravity.

The anomaly was recorded on deep space navigation (DSN) tracking data from the 
Pioneer 10 \& 11 probes \cite{Asmar05}.
An up-link radio signal is emitted from Earth at a DSN station, it is then 
received and sent back by the probe, and the down-link 
radio signal is finally received on Earth at the same or another DSN station.
{}For probes equipped with range measurement capabilities (which was not the case 
for Pioneer 10 \& 11), the ranging observable is defined as half the time elapsed 
on Earth from the emission time to the reception time.
{}For Pioneer 10 \& 11 probes, the tracking technique was based on the measurement of 
the Doppler shift, a proper observable defined as the ratio of cycle counting rates 
of reference clocks located at emission and reception stations \cite{Moyer03}.
The same information can be encoded in a Doppler velocity $\upsilon$,
with the ratio of received to emitted frequencies written as
\beqa
\label{DopplerVelocity}
&&\frac{f}{f_0}\equiv \frac{1-{\upsilon\over c}}{1+{\upsilon\over c}}
\eeqa
The observable $\upsilon$ represents a relative velocity of the probe with respect to the 
station, with relativistic and gravitational effects taken into account in the
definition (\ref{DopplerVelocity}) and perturbations due to transmission media 
effects properly accounted for \cite{Anderson02}.
  
These Doppler tracking data were analyzed during the travel of Pioneer 10 \& 11 probes
to the outer parts of the solar system.
When the probes had reached a quieter environment, after flying by Jupiter and
Saturn, a precise comparison of tracking data with predictions of GR showed
that the observed Doppler velocity departed from the calculated Doppler velocity.
The velocity was thus showing an anomaly $\delta\upsilon$ varying linearly with elapsed time
(see the Fig.~8 of Ref.~\cite{Anderson02})
\beqa
\label{anomalyP}
&&\delta\upsilon \equiv \upsilon_\mathrm{observed} - \upsilon_\mathrm{modelled} 
\simeq -a_P \left( t - t_\mathrm{in} \right) 
\eeqa
with $a_P$ an anomalous acceleration directed towards the Sun and having 
an approximately constant amplitude over a large range
of heliocentric distances (AU $\equiv$ astronomical unit)
\beqa
\label{Pioneer_acceleration}
a_P = (0.87 \pm 0.13) ~\mathrm{nm}~\mathrm{s}^{-2}\quad ,\quad 
20~\mathrm{AU}\lesssim r_P \lesssim 70~\mathrm{AU}  &&
\eeqa

It is worth emphasizing that the Pioneer anomaly has been registered on the 
two deep space probes showing the best navigation accuracy.
Other anomalous observations have been reported for Ulysses and Galileo
probes, but they were not as reliable as for Pioneer probes \cite{Anderson02}.
{}For other probes like Voyager 1 \& 2 and Cassini, the navigation accuracy was
not sufficient.
In other words, the Pioneer gravity test has been performed twice with
identical probes on similar trajectories - but escape directions opposite
in the solar system - and the same result.
This is not an impressive statistics when we compare it to the large number
of tests confirming GR.
In particular, when the possibility of an artefact onboard the probe or in its 
environment is considered, this artefact could be the same on the two probes.
However, no satisfactory explanation of this kind has been found to date, though 
intensive efforts have been devoted to this aim.
The extensive analysis of Anderson 
\textit{et al} \cite{Anderson02}, published after years of cross checks, 
has been confirmed by an independent analysis \cite{Markwardt02}.
Such independent reanalyses of the data remain an important tool to confirm
or infirm the existence of the anomaly and they now experience
a revival thanks to recently recovered data covering the whole
period of Pioneer 10 \& 11 missions from their launch to the last data 
point \cite{Turyshev05c,Nieto05earlydata,Turyshev06,Toth06}.

\section{A key question: is the Pioneer anomaly compatible %
with other gravity tests}

In this context, the question of the compatibility of the observed Pioneer 
anomaly with other gravity tests acquires the status of a key issue.
If there exist gravity theories where a Pioneer-like anomaly can take
a natural place, it is indeed of the first importance to consider these theories 
with great care because, as stated in the Introduction, the anomaly
could be the first hint of a modification of gravity at large scales,
with potentially a tremendous impact on galactic and cosmic physics.
But if there exist no such theories, the Pioneer anomaly may remain
an interesting curiosity with a potentially large impact on navigation in
the solar system, but probably lesser importance for fundamental physics.

At this point, it is worth repeating that tests of the equivalence 
principle (EP) have shown it to be preserved at a very high accuracy level, 
better than 10$^{-12}$ in laboratory experiments as well as in tracking
of the motion of Moon on its orbit around the Earth.
This is in any case a much higher accuracy than the EP violation which would 
be needed to account for the Pioneer anomaly~: the standard Newton acceleration 
at 70~UA is of the order of 1~$\mu$m~s$^{-2}$ while the Pioneer anomaly is 
of the order of 1~nm~s$^{-2}$.
Should the anomaly be interpreted in terms of an EP violation, the latter
would be of the order of 10$^{-3}$.
This does not contradict the possibility of  EP violations which are 
predicted by unification models \cite{Damour96,Damour02,Overduin00} and looked for
in space experiments with an excellent precision, such as 
MICROSCOPE \cite{MICROSCOPE} and STEP \cite{STEP}.
But such violations are expected to occur at a lower level than needed
to affect the Pioneer anomaly and we will therefore restrict our attention 
to a confrontation of GR with alternative metric theories of gravity.

In this well established metric interpretation, 
the precise form of the coupling between space-time curvature and gravity 
sources can still be discussed \cite{Weinberg72}.
Like the other fundamental interactions, gravitation may also be treated
within the framework of field theory \cite{Thirring,Feynman,Weinberg65}.
Radiative corrections due to its coupling to other fields then naturally lead 
to embed GR within a larger class of theories \cite{deWitt,Deser74,Capper74}.
Modifications are thus expected to appear \cite{Stelle,Sakharov,Adler}
in particular, though not only, at large length scales \cite{Nieto,Deffayet02,Dvali03,Gabadadze04}.
This suggests to consider GR as an effective theory of gravity valid at the length 
scales for which it has been accurately tested but not necessarily at smaller or
larger scales. 
Note that, in contrast to GR \cite{tHooft}, the fourth order theories which are a 
natural extension of GR show renormalizability 
as well as asymptotic freedom at high energies \cite{Fradkin}.
This is a strong argument for extending the gravitation theory at scales 
not already constrained by experiments, for instance 
using renormalization group trajectories \cite{Reuter}. 
Renormalizability of these theories however comes with a counterpart, that is the 
problem of ghosts, but it has been argued that this problem does not constitute a definitive 
deadend for an effective field theory valid in a limited scale domain \cite{Simon90}. 
In particular, the departure from unitarity is expected to be negligible at 
ordinary scales tested in present day universe \cite{Hawking02}.

In the following, we will briefly review the main features of a phenomenological 
framework which has been recently developed for the purpose of answering the 
question of the compatibility of the Pioneer anomaly with other gravity
tests \cite{JR05mpl,JR05cqg,JR06cqg}. 
It will be presented below as covering the whole spectrum of metric extensions 
of GR which remain in the vicinity of GR. In particular, it will be shown to 
include as particular cases the PPN extensions as well as the already evoked
modifications of Newton force law.
Let us stress that this larger family of theories is not just an {\it adhoc} 
extension showing the nice property of letting a place for the Pioneer anomaly. 
It emerges in a natural manner as the extension of GR induced by radiative 
corrections due to the coupling of gravity with other fields \cite{Jaekel95},
this idea having been explored before it was noticed that it led to 
Pioneer-like anomalies \cite{JR05mpl,JR05cqg,JR06cqg}.

\section{Post-Einsteinian metric theories of gravity}

In order to present the extensions of GR in a simple manner, we start with the linearized
version of gravitation theory \cite{JR05mpl,JR05cqg}.
We will then present some salient features of the non linear 
theory \cite{JR06cqg,JR06submitted}.

In the linearized treatment, the metric field is represented as a small perturbation 
$h_{\mu\nu}$ of Minkowski metric $\eta_{\mu\nu}$ 
\beqa
&& g _{\mu\nu} = \eta_{\mu\nu} + h_{\mu\nu} \quad,\quad
\eta_{\mu\nu} = {\rm diag}(1, -1, -1, -1)   \quad,\quad 
\left\vert h_{\mu\nu} \right\vert \ll 1 
\eeqa
The field $h_{\mu\nu}$ is a function of position $x$ in spacetime or, 
equivalently in Fourier space, of wavevector $k$.
Gauge invariant observables of the metric theory are given by curvature tensors. 
In the linearized theory, {\it i.e.} at first order in $h_{\mu\nu}$, 
Riemann, Ricci, scalar and Einstein curvatures have simple expressions 
in the momentum representation (they are given in Ref.~\cite{JR05cqg}). 

These curvature fields are similar to the gauge invariant electromagnetic fields
of electrodynamics so that, while being supported by its geometrical interpretation, 
GR shows essential similarities with other field theories \cite{Feynman,deWitt}.
This suggests that GR may be considered as the low energy effective limit
of a more complete unified theory \cite{Sakharov,Adler} which should describe
the coupling of gravity with other fields.
In any case, this theory should contain radiative corrections to the graviton propagator,
leading to a modification of the Einstein-Hilbert equation (\ref{GR_gravitation_law}) and
to a momentum dependence of the coupling between curvature and stress tensors.
In the weak field approximation, the Einstein tensor, which is divergenceless, 
has a natural decomposition on the two sectors corresponding to 
different conformal weights \cite{Jaekel95}, that is also on traceless (conformal weight 0)
and traced components (conformal weight 1). 

When considering the isotropic and stationary situation with a point-like and 
motion-less Sun of mass $M$, the general coupling between curvature and stress tensors
is thus described by two running constants $\Gts$ and $\Gtt$, which depend on the spatial 
wavevector $\mathbf{k}$ and live in the two sectors (0) and (1). 
Solutions of the extended gravitation equations (given in Refs.~\cite{JR05mpl,JR05cqg}),
anew written with spatial isotropic coordinates, depend on two potentials 
\beqa
\label{polar_metric_NP}
 g _{00} &=& 1 + 2 \PN \quad,\quad 
 g _{rr} = -(1 - 2 \PN + 2\PP) 
\eeqa
These two potentials obey Poisson equations
with running constants $\GtN$ and $\GtP$ given by
linear combinations of $\Gts$ and $\Gtt$ 
\beqa
&&-\ks^2 \PNa[\ks] =  \GtNa[\ks]{4\pi \M \over  c ^2} 
\quad,\quad a= N,P  \\
&&\GtN \equiv {4 \Gts - \Gtt \over 3} \quad,\quad 
\GtP \equiv {2 (\Gts - \Gtt) \over3}  \nonumber
\eeqa

Standard Einstein equation is recovered when the running constants $\Gts$ and  $\Gtt$
are momentum independent and equal to each other, that is also when 
\beqa
&&\stand{\GtN} \equiv  \GN \quad,\quad \stand{\GtP} = 0 \nonumber\\
&&\stand{\PN( r )} \equiv  \Pst( r ) \quad,\quad 
\stand{\PP( r )} = 0 
\eeqa
The two potentials $\PNa$ will be written as sums of these standard expressions and
anomalies which have to remain small
\beqa
\label{two_potentials}
\PNa ( r ) \equiv \stand{\PNa( r )} + \delta\PNa ( r ) \quad,\quad 
\left| \delta\PNa ( r ) \right| \ll 1 &&
\eeqa
This linearized form of the extended theory is quite useful for introducing the ideas 
in terms of an effective field theory of gravitation. It is however not sufficient
to deal with the general relation between metric and curvature tensors which involves
non linear expressions. It is no more satisfactory for the general discussion of 
gravity tests as some of them also involve non linearity of the gravitation theory.

It turns out that the extended theory may in fact be given a full non linear 
formulation, discussed in great details in Ref.~\cite{JR06cqg}.
Most formulas are thus written more conveniently in terms of 
Schwartzschild coordinates \cite{MTW} 
\beqa
\label{Schwartzschild_isotropic_metric}
&&d\s^2 = \gb_{00}(\lr) c ^2 d t ^2 + \gb_{ r  r }(\lr)d\lr^2 
- \lr^2 \left(d\theta^2 + {\rm{sin}}^2\theta d\varphi^2\right)\nonumber\\
&&\gb_{\mu\nu}( r ) \equiv \stand{\gb_{\mu\nu}( r )} + \delta\gb_{\mu\nu} ( r )  \quad,\quad 
\left| \delta\gb_{\mu\nu}( r ) \right| \ll 1 
\eeqa
with the standard GR solution treated exactly 
\beqa
\label{metric_st}
&&\stand{\gb_{00}} = 1-2{\kappa\over\lr} =  -{1\over\stand{\gb_{ r  r }}} 
\eeqa
and the anomalous metric dealt with at first order. 
It is possible to define in the non linear theory
two potentials $\PbN$ and $\PbP$ which generalize (\ref{two_potentials})
while taking into account the non linear corrections involving
powers of $\kappa/\lr$.

We do not reproduce here the corresponding calculations but emphasize a few
salient features of the results. First, the phenomenological freedom
of the extended framework is represented by the two functions 
$\delta\gb_{00}(r)$ and $\delta\gb_{rr}(r)$ which contain the same information,
through the appropriate transformations \cite{JR06cqg},
as $\delta g_{00}(r)$ and $\delta g_{rr}(r)$, or $\delta\PN(r)$ and $\delta\PP(r)$,
or $\delta\GN[\ks]$ and $\delta\GP[\ks]$.
They can as well be described by Einstein curvatures 
$E ^0_0$ and $E ^r_r $ which no longer vanish outside the source \cite{JR06cqg}.
The PPN family is recovered as a particular case which already shows an 
anomalous behaviour of Einstein curvatures non null apart from the gravity source. 
Anew, the post-Einsteinian metric theory is nothing but an 
extension of this anomalous behaviour with more general dependences of the curvatures 
versus the distance $r$ to the Sun.
In loose words, the post-Einsteinian metric theories can be 
thought of as an extension of PPN metric with PPN parameters
no longer constants but now functions of $r$.

\section{Phenomenological consequences}

The new phenomenological framework is characterized by the two functions 
$\delta g_{00}(r)$ and $\delta\left(g_{00} g_{rr}\right)(r)$. 
The first function represents an anomaly of the Newton potential which
has to remain small to preserve the good agreement between GR and 
gravity tests performed on planetary orbits \cite{JR04,Coy03}.
Meanwhile, the second sector represents an extension of PPN phenomenology 
with a scale dependent Eddington parameter $\gamma$. 
It opens an additional phenomenological freedom with respect to the mere 
modification of the Newton potential and this freedom opens the possibility 
to accomodate a Pioneer-like anomaly besides other gravity 
tests \cite{JR05mpl,JR06cqg}.

Recent publications force us to be more specific on the relation between the Pioneer 
anomaly and modifications of the Newton potential, \textit{i.e.} anomalies in the first sector
according to the terminology of the preceding paragraph. Interpreting the Pioneer
anomaly in such a manner requires that $\delta g_{00}$ varies roughly as $r$
at the large radii explored by Pioneer probes. If this dependence also holds at smaller 
radii \cite{Anderson02}, or if the anomaly follows a simple Yukawa law \cite{JR04},
one deduces that it cannot have escaped detection in the more constraining tests 
performed with martian probes \cite{Reasenberg79,Anderson96}.
Brownstein and Moffat have explored the possibility that the linear dependence 
holds at distances explored by Pioneer probes while being cut 
at the orbital radii of Mars \cite{Brownstein06}.
Other authors \cite{Iorio06,Tangen06} have in contrast argued that the ephemeris of 
outer planets were accurate enough to discard the presence of the required linear 
dependence in the range of distances explored by the Pioneer probes. 
This argument has been contested by the authors of Ref.~\cite{Brownstein06}
and the conflict remains to be settled. 

The authors of Refs.~\cite{Iorio06,Tangen06} have pushed their claim 
one step farther by restating their argument as an objection to the very possibility 
of accounting for the Pioneer anomaly in any viable metric theory of gravity. 
This claim is clearly untenable because it only considers metric anomalies
in the first sector while disregarding those in the second sector. 
At this point, we want to repeat that the discussion of the compatibility
of metric anomalies with observations performed in the solar system 
has to be done carefully, accounting for the presence of the two sectors
as well as for possible scale dependences.
This question has already been discussed in Refs.~\cite{JR05cqg,JR06cqg} 
for the cases of deflection experiments on electromagnetic sources passing 
behind the Sun \cite{Shapiro04,Bertotti03}. 
It has a particularly critical character for the ranging experiments 
which involve directly the Shapiro time delay \cite{Shapiro99}.

The second potential $\delta\left(g_{00} g_{rr}\right)(r)$ naturally produces an
anomaly on Doppler tracking of probes with escape trajectories in the outer solar
system. This Pioneer-like anomaly can be calculated
by taking into account the perturbations on probe motions as well as on
light propagation between stations on Earth and probes.
The time derivative of the Doppler velocity thus computed can be written 
as a Doppler acceleration $a$ and the anomaly evaluated as the difference
of the values obtained in the extended and standard theories
\beqa
\label{Pioneers_acceleration}
\delta a \equiv a_\mathrm{extended} - a_\mathrm{standard} \quad,\quad
a\equiv \frac{d\upsilon}{dt} 
\eeqa
The result of the calculation given in Refs.~\cite{JR05mpl,JR05cqg} was
unfortunately corrupted by a mistake.
The mistake has been corrected in a recent publication 
Ref.~\cite{JR06submitted} which also contains the evaluation of an 
annually modulated anomaly coming out, as the secular anomaly, as a 
natural consequence of the presence of an anomalous metric in the second sector. 
As observations of such annual anomalies are reported in Ref.~\cite{Anderson02},
this situation certainly pleads for pushing this study farther and comparing the 
theoretical expectations with the newly recovered Pioneer 
data \cite{Turyshev05c,Nieto05earlydata,Turyshev06,Toth06}.

More generally, these data will make available a lot of informations
on the status of the probes as well as on the Doppler tracking details,
for the whole duration of Pioneer 10 \& 11 missions 
from their launch to the last data points.
The recovery is now completed at JPL \cite{Turyshev06data} and the upcoming
data analysis planned as an international effort \cite{ISSI}.
Numerous open questions can potentially be solved by this new analysis.
The systematics can certainly be much better controlled while
several important properties of the force - direction, time variation
of the secular anomaly, annual or diurnal modulations, spin contribution,
$\ldots$ - can be more precisely characterized.
Then, the availability of early data may make possible to confirm whether
or not the anomaly arises at Pioneer 11 at Saturn encounter,
as it is suggested by Fig.~7 of Ref.~\cite{Anderson02}.
{}Finally, the data will be confronted to the detailed predictions now available
for a variety of theoretical proposals.

If we follow the line of thought presented in this paper, the confrontation
of data with extended metric theories of gravitation is of particular interest,
as the anomaly observed on the trajectories of the Pioneer 10 \& 11 probes may well be
a first hint of a modification of gravity law in the outer part of the solar system. 
This possibility would have such a large impact on fundamental physics, astrophysics
and cosmology that it certainly deserves further investigations.
The evaluations presented in Ref.~\cite{JR06submitted} will allow one to address
these questions in a well defined theoretical framework.
It is only after a quantitative comparison, taking into account the details
known to be important for data analysis \cite{Anderson02}, that it will be 
possible to know whether the post-Einsteinian phenomenological 
framework shows the capability of fitting the Pioneer observations. 

When using the corrected expression for the secular anomaly,
identification with the observed Pioneer anomaly now points to a 
quadratic dependence of the second potential with radius. 
This corresponds to a constant curvature 
(see the evaluations in Ref.~\cite{JR06cqg})
with an unexpectedly large value in the outer solar system 
(see Ref.~\cite{JR06submitted}).
This quadratic dependence may have to be cut off at distances exceeding
the size of the solar system as well as in the inner solar system
in order to pass Shapiro tests on martian probes.
As already stressed, it is of crucial importance to check out
that the modification of GR needed to produce the Pioneer anomaly 
does not spoil its agreement with other gravity tests.
At the same time, this study can lead to Pioneer-related anomalies,
produced by the same metric anomalies, but to be looked for in other
kinds of experiments or, in some cases, by having a new look
at data of already performed experiments.

The second potential $\delta\PP$ has a direct effect on the
propagation of light rays.
It affects the Eddington deflection experiments as well as the
ranging experiments which are sensitive to the Shapiro time delay.
These experiments can in fact be described as determining the Eddington 
parameter $\gamma$, with the new feature that the latter can now
depend on the heliocentric distance (more discussions in
Ref.~\cite{JR05cqg} for deflections amplified near occultation).
The results are reduced to PPN ones when $\gamma$ is a constant.
Otherwise, they show that deflection or ranging tests can reveal the presence 
of $\delta\PP$ in the vicinity of the Sun
through a space dependence of the parameter $\gamma$.
Such a potential dependence might already be looked for through a reanalysis 
of existing data, such as VLBI measurements \cite{Shapiro04}, Cassini 
experiment \cite{Bertotti03}, or HIPPARCOS data \cite{Hipparcos}.
It may also be studied in the future through higher accuracy Eddington tests,
made possible by the global mapping of deflection over the sky in the GAIA 
project \cite{Gaia}, or by the high accuracy LATOR mission \cite{LATOR05}. 
{}For this kind of tests, the goal can be described as a construction of the 
dependence of the deflection versus the elongation of the ray with respect 
to the Sun. This function directly probes the space dependence of the second 
potential \cite{JR05cqg} and its unambiguous experimental determination
will either produce a clear signature of a deviation of GR
or put improved constraints on the existence of the second potential
at heliocentric distances smaller than 1 AU.

The presence of $\delta\PP$ can also be sensed in planetary tests. 
In particular, the perihelion precession of planets has been evaluated 
in the non linear theory \cite{JR06cqg}.
The expression there given, written as an anomaly with 
respect to GR and truncated after leading ($\propto e^0$) and sub-leading 
($\propto e^2$) orders in the eccentricity $e$ of the planetary orbit,
shows that the perihelion precession can be used as a sensitive probe
of the value and variation of the second potential.
Note that the second potential could in principle be present at the long
distances explored by the Pioneer probes, but not at the smaller distance 
corresponding to the radius of Mars orbit.
This entails that it would be extremely interesting to track with 
accuracy the motions of small bodies which may have significant radial
velocities while being at large heliocentric distances.
This possibility of testing GR by following small bodies can be considered 
as a further fundamental challenge for GAIA \cite{Gaia}. 

Generally speaking, the eccentricity of the orbits plays a key role in 
Pioneer-related anomalies. It takes large values for Pioneer-like probes
which sense $\delta\PP$ whereas it is zero for circular orbits which do not.
This suggests to devote a dedicated analysis to the intermediate situation,
not only for the two categories of bound and unbound orbits, but also for the 
flybies used to bring Pioneer-like probes from the former category to 
the latter one.
It would be worth studying planetary probes on elliptical
orbits, for example on transfer orbits from Earth to Mars or Jupiter.
Another natural target for such a study could be LISA with its three crafts
on slightly elliptical orbits \cite{Lisa}. 

{}Finally, there are strong motivations for new missions designed to study the anomaly 
and try to understand its origin \cite{PAEM05}.
A cheaper and quicker alternative could be to fly dedicated passenger instruments 
on planetary missions with different primary purposes.
In the meantime, a wise strategy is to develop and validate enabling technologies,
such as laser and radio techniques for ranging, accelerometers for controlling the deviation
from geodesic motion, accurate clocks on board for measuring separately the two components
of the metric.

\section*{Acknowledgements}

Thanks are due for discussions to the members of the ``Deep Space Gravity Probe'' team
(H.~Dittus \textit{et al}) \cite{PAEM05}, of the ``Pioneer Anomaly Investigation Team'' 
(S.G.~Turushev \textit{et al}) \cite{ISSI}, and of the ``Groupe Anomalie Pioneer'',
in particular F.~Bondu, P.~Bouyer, B.~Christophe, J.-M.~Courty, B.~Foulon, S.~L\'{e}on, 
A.~L\'{e}vy, G.~Metris and P.~Touboul.

\def\etal{\textit{et al }}
\def\ibid{\textit{ibidem }}
\def\url#1{{\rm #1}}
\def\arxiv#1{{\rm #1}}
\def\REVIEW#1#2#3#4{\textit{#1} \textbf{#2} {#4} ({#3})}
\def\BOOK#1#2#3#4{\textit{#1} ({#2}, {#3}, {#4})}
\def\BOOKed#1#2#3#4#5{\textit{#1}, #2 ({#3}, {#4}, {#5})}
\def\Name#1#2{#2 #1}

\end{document}